\documentclass[aps,prd,twocolumn,
comment, 
superscriptaddress,groupedaddress,
amsmath,amssymb,floatfix,nofootinbib]
{revtex4}  
\usepackage{float}
\usepackage{graphicx}
\usepackage{bm}
\usepackage[usenames,dvipsnames]{color}
\usepackage{slashed}
\usepackage{hyperref}
\usepackage{slashed}

\begin{document}

%

\let\a=\alpha      \let\b=\beta       \let\c=\chi        \let\d=\delta
\let\e=\varepsilon \let\f=\varphi     \let\g=\gamma      \let\h=\eta
\let\k=\kappa      \let\l=\lambda     \let\m=\mu
\let\o=\omega      \let\r=\varrho     \let\s=\sigma
\let\t=\tau        \let\th=\vartheta  \let\y=\upsilon    \let\x=\xi
\let\z=\zeta       \let\io=\iota      \let\vp=\varpi     \let\ro=\rho
\let\ph=\phi       \let\ep=\epsilon   \let\te=\theta
\let\n=\nu
\let\D=\Delta   \let\F=\Phi    \let\G=\Gamma  \let\L=\Lambda
\let\O=\Omega   \let\P=\Pi     \let\Ps=\Psi   \let\Si=\Sigma
\let\Th=\Theta  \let\X=\Xi     \let\Y=\Upsilon
%


\def\cA{{\cal A}}                \def\cB{{\cal B}}
\def\cC{{\cal C}}                \def\cD{{\cal D}}
\def\cE{{\cal E}}                \def\cF{{\cal F}}
\def\cG{{\cal G}}                \def\cH{{\cal H}}
\def\cI{{\cal I}}                \def\cJ{{\cal J}}
\def\cK{{\cal K}}                \def\cL{{\cal L}}
\def\cM{{\cal M}}                \def\cN{{\cal N}}
\def\cO{{\cal O}}                \def\cP{{\cal P}}
\def\cQ{{\cal Q}}                \def\cR{{\cal R}}
\def\cS{{\cal S}}                \def\cT{{\cal T}}
\def\cU{{\cal U}}                \def\cV{{\cal V}}
\def\cW{{\cal W}}                \def\cX{{\cal X}}
\def\cY{{\cal Y}}                \def\cZ{{\cal Z}}


\def\be{\begin{equation}}
\def\ee{\end{equation}}
\def\bea{\begin{eqnarray}}
\def\eea{\end{eqnarray}}
\def\bm{\begin{matrix}}
\def\em{\end{matrix}}
\def\bpm{\begin{pmatrix}}
    \def\epm{\end{pmatrix}}

{\newcommand{\lsim}{\mbox{\raisebox{-.6ex}{~$\stackrel{<}{\sim}$~}}}
{\newcommand{\gsim}{\mbox{\raisebox{-.6ex}{~$\stackrel{>}{\sim}$~}}}
\def\mpl{M_{\rm {Pl}}}
\def\gev{{\rm \,Ge\kern-0.125em V}}
\def\tev{{\rm \,Te\kern-0.125em V}}
\def\mev{{\rm \,Me\kern-0.125em V}}
\def\ev{\,{\rm eV}}

\title{\boldmath  Moduli induced cogenesis of baryon asymmetry and dark matter}
\author{Mansi Dhuria}
\email{mansi@prl.res.in}
\affiliation{Physical Research Laboratory, Navrangpura, Ahmedabad 380 009, India}
\author{Chandan Hati}
\email{chandan@prl.res.in} 
\affiliation{Physical Research Laboratory, Navrangpura, Ahmedabad 380 009, India}
\affiliation{Indian Institute of Technology Gandhinagar, Chandkheda, Ahmedabad 382 424, India.}
\author{Utpal Sarkar}
\email{utpal@prl.res.in} 
\affiliation{Physical Research Laboratory, Navrangpura, Ahmedabad 380 009, India}

\begin{abstract}  
We study a cogenesis mechanism in which the observed baryon asymmetry of the universe and the dark matter abundance can be produced simultaneously at low reheating temperature without violating baryon number in the fundamental vertex. In particular, we consider a model which could be realized in the context of type IIB large volume string compactifications. The matter superfields in this model include additional pairs of color triplet and singlet superfields in addition to the Minimal Supersymmetric Standard Model (MSSM) superfields. Assuming that the mass of the additional singlet fermions is ${\cal O}(\gev)$ and color triplet fermions is ${\cal O}(\tev)$, we show that the modulus dominantly decays into the additional color triplet superfields. After soft supersymmetry (SUSY) breaking, the lightest eigenstate of scalar component of color triplet superfield further decays into fermionic component of  singlet superfield and quarks without violating baryon number. Assuming R-parity conservation, it follows that the singlet superfield will not further decay into the SM particles and therefore it can be considered as a stable asymmetric dark matter (ADM) component. We find that the decay of the lightest eigenstate of scalar component of color triplet superfield gives the observed baryon asymmetry in the visible sector, an asymmetric dark matter component with the right abundance and naturally explains cosmic coincidence.

\end{abstract}
\maketitle
 

\section{Introduction}
\label{Introduction} 
In cosmology some of the important puzzles are related to the origin of baryon asymmetry of the universe and the nature of dark matter. The comparable values of dark matter density and baryon density \cite{Ade:2015xua} ${\Omega_{DM} h^{2}_{0}} \sim 5~{\Omega_{B}h^{2}_{0}}$, points to the possibility that they might have a common origin. However, the standard paradigm adopts completely different mechanisms  to explain observable baryon asymmetry of the universe and dark matter abundance. The baryon asymmetry is generated from an initially baryon-antibaryon symmetric universe by considering baryon number (B), C and CP  violating processes that went out of equilibrium in the early universe, while the dark matter density is  produced by considering  weakly interacting massive particles (WIMPs) (with mass around ${\cal O}(100) \gev$) with the relic density being determined by the freeze out condition. The fact that they have a comparable abundance is often referred to as the ``cosmic coincidence" puzzle. Recently, the CDMS collaboration has reported an excess in the dark matter events \cite{Agnese:2013cvt} which sets an upper limit of ${\cal O}(10^{-41})~{\rm cm}^2$ on the value of spin-independent (SI) dark matter-nucleon cross section for dark matter mass around ${10} \gev$ at 3.1$\sigma$ significance level. The excess reported by the CoGeNT collaboration \cite{Aalseth:2012if} also hints at a light dark matter mass, almost in the same region of parameter space. The data taken by the XENON100 experiment \cite{Aprile:2012nq} also gives a very stringent constraint on SI dark matter-nucleon cross section which points towards a dark matter mass around ${\cal O}(\gev)$. The light dark matter is often also motivated due to the possibility of  explaining 3.5 KeV X-ray line by radiative decay of ${\cal O}(\gev)$ neutral dark matter particle \cite{Allahverdi:2014dqa}. However, for an ${\cal O}(\gev)$ mass the thermal WIMPs give over-abundance of dark matter particle for annihilation cross-section less than $10^{-26}$ ${\rm cm}^2$, and thus the alternative schemes where an ${\cal O}(\gev)$ mass dark matter can be accommodated have gained significant attention. To this end, the cogenesis scenarios are particularly interesting because they have an attractive feature of explaining the observed baryon asymmetry of the universe together with an asymmetric dark matter component which can naturally satisfy the criterion for ${\cal O}(\gev)$ mass dark matter. Furthermore, the apparent coincidence of the baryon and dark matter densities can also be addressed in such a framework using the underlying connection between the baryogenesis scenarios and dark matter production. There exists several different mechanisms in the literature  \cite{Gu:2009hj}, which address simultaneous generation of baryon (or lepton) asymmetry and the asymmetric dark matter abundance. The cogenesis of both without violating $B$ or $B-L$ is discussed in Refs. \cite{Davoudiasl:2010am}.
 
From a top-down model building perspective, an UV completed supersymmetric model may entail the existence of WIMPs, known as the modulus (moduli). In ${\cal N}=1$ supergravity limit of string theory, the moduli  appear while compactification of the extra dimensions takes place \cite{Greene:1996cy}. The decay of moduli fields have significant implications for the cosmological history of the universe \cite{Kane:2015jia}. The entropy released due to late decay of the lightest modulus dilutes the existing baryon asymmetry of the universe as well as the relic abundance of dark matter produced at high scale. However, the correct amount of dark matter can be produced non-thermally from the decay of modulus into the lightest supersymmetric particle. The non-thermal realization of dark matter is discussed in Refs. \cite{Acharya:2008bk,Allahverdi:2013noa,Baer:2014eja,Blinov:2014nla} in the context of Minimal Supersymmetric Standard Model (MSSM) and string-motivated models. Given that the decay of heavy modulus leads to very low reheating temperature, it renders electroweak baryogenesis and leptogenesis impossible. However, it is possible to accommodate direct baryogenesis and correct dark matter abundance by considering late-decaying moduli in the schemes implementing presence of additional color triplet superfields along with MSSM superfields \cite{Allahverdi:2010im,Allahverdi:2010rh,Allahverdi:2013tca}, and with implementation of other mechanisms \cite{Kitano:2008tk,Ishiwata:2014cra}. The coincidence problem has also been addressed by considering Affleck Dine (AD) baryogenesis in the presence of moduli in Refs. \cite{Kawasaki:2007yy,Kane:2011ih}. Interestingly, the decay of lightest string modulus into its superpartner axion can also explains the source of dark radiation \cite{Cicoli:2012aq,Higaki:2012ar,Allahverdi:2014ppa}. 

In this work, we propose a model for moduli induced cogenesis which simultaneously generates the baryon asymmetry of the universe and an asymmetric dark matter (ADM) component with a dark matter mass around $5 \gev$. In this model, the particle content include two additional iso-singlet color triplet superfields ${\chi}$ and  ${\bar \chi}$ with hypercharges $-4/3$ and $4/3$ respectively and two singlet superfields ${\cal N}$ and ${\bar {\cal N}}$ \footnote{Though the particle content is quite similar to the model considered in Ref. \cite{Allahverdi:2013tca}, the cogenesis mechanism producing baryon asymmetry and dark matter discussed in this work is completely different.}. Due to the presence of the pair of the additional superfields with opposite hypercharges, the terms analogous to the Giudice-Masiero term \cite{Giudice:1988yz} in the K\"{a}hler potential dictate the decay width of modulus into both colored and singlet superfields. In ${\cal N}=1$ supergravity,  the effective supersymmetric mass terms as well as soft SUSY breaking terms depend on coupling strength of the hidden sector field (modulus) to the visible sector fields \cite{Brignole:1997dp}. Interestingly, the effective masses of additional colored(singlet) superfields are also governed by the same Giudice-Masiero like term(s) considered in the k\"{a}hler potential. Therefore, the coefficient of interaction term responsible for the decay of modulus into the pair of colored(singlet) superfields i.e. the coefficient of new Giudice-Masiero like term(s) can be constrained  based on given masses of the superfields. In this model the mass of color triplet superfields being heavier as compared to the mass of singlet superfields, the modulus would preferably decay into pair of color triplet superfields. Now the scalar component of color triplet superfields further decay into quarks and additional singlet fermions and the baryon number of the color triplet superfield gets distributed between quark and additional singlet fermion. The conservation of R-parity ensures that the singlet fermion will further not decay into the Standard Model (SM) particles and therefore can be considered as dark matter component. The decay process is baryon number conserving at tree level as well as at one-loop level, however, it is CP asymmetric at one-loop level due to the presence of soft SUSY breaking terms. Consequently, an asymmetry is generated in both the visible and the dark matter sector. We find that the symmetric component of dark matter gets annihilated for a dark matter mass ${\cal O}(\gev)$, and the required order of baryon asymmetry and dark matter relic abundance can be successfully generated in this mechanism for certain values of Yukawa couplings.

The outline of the rest of the paper is as follows. In section {\bf II}, we briefly describe a phenomenological model that can be obtained as a low energy limit of the large volume scenario (LVS) proposed in the context of type IIB string compactifications. In section {\bf IIA}, we discuss all possible decay modes of the modulus and give the corresponding decay widths showing that the modulus dominantly decays into pair of Higgs, axions and additional color triplet and singlet superfields introduced in the low energy spectrum, depending on the coefficients of the interaction terms. In section {\bf II B}, we show that due to the heavier mass of color triplet superfields as compared to singlet superfields,  the modulus preferably decays into pair of color triplet superfields. In section {\bf III}, we propose a new mechanism which can simultaneously generate the observed baryon asymmetry and the dark matter abundance without violating baryon number in the fundamental vertex.  In section {\bf III A}, we discuss the possible decay mode of scalar component of color triplet superfield into quarks and additional singlet fermions, generating a baryon asymmetry. Next we argue that in a R-parity conserving scenario, the singlet fermion will further not decay into the (MS)SM particles and therefore can be considered as a stable dark matter candidate. In section {\bf III B}, we discuss a mechanism to annihilate the symmetric component of dark matter, leaving only the asymmetric component accounting for the dark matter relic abundance. In section {\bf IV}, we summarize our results and conclude.
 \section{A Phenomenological Model Based on Large Volume Scenario and Modulus decay}
 
 The presence of  gravitationally coupled moduli fields can have significant impact on the standard cosmology. During inflation i.e. when the Hubble expansion rate $H_{\rm inf} >> m_{\Phi}$, modulus ($\Phi$) gets significantly displaced from the minimum of its potential \cite{Dine:1995kz}. Thus, if one takes into consideration the presence of modulus and high scale inflation, it is a rather generic consequence to expect the modulus to be displaced from the low-energy  minimum by an amount $ \left|\Delta \Phi \right| = \left| {\langle \Phi \rangle}_{\rm inf} - {\langle \Phi \rangle}_{0}  \right | \approx M_{P}$. Since the energy density of these oscillations dilutes in the same way as non-relativistic matter, they will come to dominate the expansion of the universe. This will continue until the modulus decays at a time $t\sim \Gamma^{-1}_{\Phi}$,  transferring the remaining
oscillation energy into radiation, hence reheating the universe at a late time. The reheating temperature after the modulus decay is given by  \cite{Kane:2015jia}, 
\begin{equation}
T_R=\frac{1}{{g_*}^{1/4}} \sqrt{\Gamma_{\Phi} M_P},
\end{equation}
 where ${g_*}$ is the total number of relativistic degrees of freedom and $\Gamma_{\Phi}$ is the decay width of the modulus.
The number density of any particle $X$ produced from the decay of the modulus is given by
\begin{equation}
\label{eq:YX}
Y_{X}= Y_{\Phi} {\rm Br}_{X}= \frac{ 3 T_R}{4 M_\Phi} {\rm Br}_{X}
\end{equation}

It has been a challenging task to obtain a realistic low energy spectrum in string compactifications. The foremost step while constructing reals in string compactifications is the issue of moduli stabilization. A realistic model should be able to realize de-Sitter minima and also avoid Cosmological Moduli Problem (CMP) \cite{Banks:1993en}.  The Large Volume Scenario (LVS) has been considered as an ideal framework to build consistent MSSM-(like) chiral model in which  soft terms are calculated explicitly.  Let us consider the Large Volume type IIB compactification scheme initially proposed in \cite{{Balasubramanian:2005zx},Cremades:2005ir}. The volume of the Calabi-Yau (CY) manifold is of swiss-cheese type, given by ${\cal V} \sim ( \tau^{3/2}_B -  \lambda_\alpha \tau^{3/2}_{\alpha})$, where $\tau_B$ denotes the big divisor volume modulus which mainly controls the size of the CY volume and $\tau_{\alpha}$'s correspond to small divisor or blow-up moduli fields. The volume moduli are complexified by associating them with four-form axions. The spectrum also takes into consideration dilaton (S), complex structure moduli (U) and two-form axions. The K\"{ahler} potential of the effective theory includes k\"{a}hler moduli and perturbative $\alpha^{\prime}$ corrections. The superpotential includes nonperturbative contribution effects on the small blow-up mode. Interestingly, the (non)perturbative corrections in the effective potential lead to the nonsupersymmetric anti-di Sitter minima when the volume of the Calabi-Yau manifold is very large  \cite{Balasubramanian:2005zx}. The visible/MSSM sector in this scenario is realized  by including D3/D7 branes on blow-up modes \cite{Conlon:2005ki}. The soft SUSY breaking terms are calculated from the nonvanishing F-terms corresponding to the hidden-sector moduli. The pattern of the soft-terms depend on the location of the D-brane in the bulk geometry. If the MSSM-like divisor is placed in the proximity of main source of SUSY breaking sector, it will give soft term masses of the order of gravitino mass. If location of D-brane is geometrically separated from dominant SUSY breaking sector, it will give smaller mass of sparticles as compared to the gravitino mass, known as sequestered string models.  

In this work, we follow the sequestered Large Volume Models discussed in Refs. \cite{Blumenhagen:2009gk,Aparicio:2014wxa}. The relevant scales in this model are given by
\begin{enumerate}
\item[(i)]
  string scale $M_{\rm string}= M_{P}/ {\sqrt{\cal V}}$,
   \item[(ii)]
   { Kaluza-Klein~scale} $M_{KK}= M_{P}/ {{\cal V}^{\frac{2}{3}}}$,
  \item[(iii)]
  { gravitino mass} $m_{3/2}= {W_0 M_P/{\cal V}}$,
    \item[(iv)]
     { lightest ``big" divisor modulus} $M_{\tau_B}= m_{3/2}/{\sqrt{\cal V}}$.
\end{enumerate}
The soft terms as well as supersymmetric mass terms are evaluated by expanding the K\"{a}hler potential and superpotential as a power series expansion in the matter superfields respectively. The analysis of soft terms in Ref. \cite{Aparicio:2014wxa} is given in two limits, namely (i) local limit (ii) ultra-local limit. The classification is based on the precise form of the K\"{a}hler metric which is used to obtain Yukawa couplings  independent of the compactification volume. Since the analysis of the soft terms also depends on that particular form of K\"{a}hler metric, it  generates different pattern of soft SUSY breaking terms in different limits.  In this paper, we consider ultra local limit in which 

\label{eq:scales}
\bea
  (\rm i)&&{\rm gaugino~mass~}\; m_{1/2}   \sim m_{3/2}/ {\cal V}, \nonumber\\
  (\rm ii)&&{\rm sfermion~mass~}\; m^{2}_{\alpha \beta}/m^{2}_{\rm soft} \sim {\cal O}(m^{2}_{1/2})\nonumber\\
  (\rm iii)&&{\rm higgsino~mass~parameter~}\; \mu  \sim {\cal O}(m_{1/2}),\nonumber\\
  (\rm iv)&&{\rm soft~Higgs~mixing~term~}\; B{\mu}  \sim  {\cal O}(m^{2}_{1/2}),\nonumber\\
  (\rm v)&&{\rm trilinear~soft~terms~}\;  {\cal A}_{\alpha \beta \gamma} \sim {\cal O}(m_{1/2}).
\eea
 The requirement of $m_{\rm soft} \sim {\cal O}(\tev)$ constrains the value of CY volume ${\cal V} \sim {\cal O}(1) \times 10^7$ in string length units. The choice of ${\cal V} \sim 10^{7}$ also provides 60 $e$-folds of inflation, generating 
right amount of density perturbations in this model \cite{Burgess:2013sla}.

We assume that the blow-up mode upon which visible sector is realized by wrapping D3/D7-brane, includes aforementioned new set of (s)particles in addition to the MSSM spectrum. Therefore, the matter K\"{a}hler potential for our model includes the soft SUSY breaking terms corresponding to additional superfields also.
\subsection{Modulus decay}
There are multiple moduli present in the LVS model described in Refs. \cite{Blumenhagen:2009gk,Aparicio:2014wxa}, however only the lightest modulus couples to the (MS)SM as well as the additional fields.  Below we give the kinematically possible decay modes of the lightest ``big" divisor volume modulus into different modes, studied extensively in Ref. \cite{Cicoli:2012aq}.
  
 The K\"{a}hler potential  involving the ``big" divisor volume is given by
  \begin{equation}
  K= -3 \  ln \left(T_B + {\bar T_B} \right),
  \end{equation}
with $T_B= \tau_B + i a_B$, where $\tau_B$ is the real volume modulus and $a_B$ is four-form axion. This leads to the interaction terms
  \begin{equation}
  {\cal L}= \frac{3}{ 4 \tau^{2}_{B}} \partial_{\mu} {\tau_B} \partial^{\mu} {\tau_B} + \frac{3}{ 4 \tau^{2}_{B}} \partial_{\mu} {a_B} \partial^{\mu} {a_B}.
  \end{equation}
Canonically normalizing the volume modulus $\Phi= \sqrt{\frac{3}{2}} \ln {\tau_B}$, one obtains
  \begin{equation}
  {\cal L}= \frac{1}{2} \partial_{\mu} \Phi \partial^{\mu} \Phi +\frac{1}{2} \left(\frac{3}{2} {\rm exp}\left[-2 \sqrt{\frac{2}{3}} \Phi \right] \right) \partial_{\mu} a_B \partial^{\mu} a_B.
  \end{equation}
  (i) Decay into axions: Utilizing the above, the decay width for the modulus decaying into axions is given by
  \begin{equation}
\Gamma_{\Phi \rightarrow a_B a_B} \sim \frac{1}{48 \pi}  \frac{m^{3}_\Phi }{M^{2}_{P}}.
  \end{equation}
  (ii) Decay into gauge bosons:  The coupling of the modulus to the gauge bosons is obtained through gauge kinetic function. In LVS model, the SM arises from wrapping D-branes on the blown-up modulus, so there is no direct tree-level coupling of the volume modulus ${\tau}_B$ to gauge bosons ($A^{\mu}$).  The effective interaction term appearing at one-loop level is given by
  \begin{equation}
  {\cal L}= \frac{\lambda_{a} \alpha_{\rm SM}}{4 \pi} \Phi F_{\mu \nu} F^{\mu \nu} \ +...\; .
  \end{equation}
This leads to decay width given by
  \begin{equation}
  \Gamma_{\Phi \rightarrow A^{\mu} A^{\mu}}\sim \left(\frac{\alpha_{\rm SM}}{4 \pi}\right)^2 \frac{m^{3}_{\Phi}}{M^{2}_{P}} \; .
  \end{equation}  
  (iii) Decay into MSSM Scalars: The couplings to matter scalars are given by
\begin{equation}
\label{K}
K= -3 \  ln \left(T_B + {\bar T_B} \right) + \frac{{\cal C} {\bar {\cal C}}}{T_B +{\bar T_B}} \; .
\end{equation}
After canonical normalization, the interaction term is  given by
\begin{equation}
{\cal L}= \frac{1}{2} \sqrt{\frac{2}{3}} \left ({\bar {\cal C}} \Box {\cal C} + {\cal C} \Box  {\bar {\cal C}} \right).
\end{equation}
The decay width for the matter scalars is given by
\begin{equation}
\label{decysc}
\Gamma_{\Phi \rightarrow {\cal C} {\bar  {\cal C}}} \sim \frac{m^{2}_{\rm soft} m_{\Phi}}{M^{2}_{P}}.
\end{equation}
(iv) Decay into matter fermions and gauginos:
Using equation (\ref{K}), the coupling of matter fermions and gauginos to the volume modulus is given by
\begin{equation} 
{\cal L}= {\lambda} \frac{\Phi}{M_P} {\bar \lambda_{\mu}} {\bar \sigma}^m D_{m} \lambda^{\mu},
\end{equation}
It leads to a decay width
\begin{equation}
\label{decyferm}
\Gamma_{\Phi \rightarrow \lambda {\bar \lambda}} \sim \frac{m^{2}_{\lambda} m_{\Phi}}{M^{2}_{P}}.
\end{equation}
(v) Decay into the Higgs:
The coupling of the modulus to the Higgs fields is dominated by the Guidice-Maisero coupling in the K\"{a}hler potential
\begin{eqnarray}
&& K= -3  \ ln \left(T_B + {\bar T_B}\right) +\frac{1}{ \left(T_B + {\bar T_B}\right)}\left(H_u {H_u}^{\dagger} +H_d {H_d}^{\dagger} \right) \nonumber\\
&& + \left( \frac{ z H_u  H_d} {T_B + {\bar T_B}} + h.c. \right) \; .
\end{eqnarray}
After canonical normalization, the decay width is given by
\begin{equation}
\Gamma_{\Phi \rightarrow H_u H_d }\sim \frac{z^2}{48 \pi} \frac{m^{3}_\Phi }{M^{2}_{P}}.
\end{equation}
If the Higgs possess a shift symmetry \cite{Hebecker:2012qp},  then $z=1$ and decay width is proportional to ${m^{3}_{\Phi}}/{M^{2}_{P}}$. \\

(vi) Decay into color triplets:  
We consider an analogue of Guidice-Maisero term for additional color triplets superfields under consideration.  So, the couplings of the modulus to the pair of color triplets is  given by
\begin{equation}
K=   \frac{1}{{T_B+ {\bar T}_B}}\left({\chi}{\chi}^{\dagger} + {\bar \chi} {\bar \chi}^{\dagger} \right)+  \left( \frac{z_{\chi} {\chi} {\bar \chi}}{{T_B+ {\bar T}_B}} + h.c.\right),
\end{equation}
where $z_{\chi}$ is an undetermined constant. After canonical normalization of the volume modulus, the interaction term is given by
\begin{equation}
{\cal L}= \frac{1}{2} \sqrt{\frac{2}{3}} \left ({\partial}^{2}  \Phi^{\dagger} \right) {\chi} {\bar \chi}.
\end{equation}
The decay width for color triplets is given by
\begin{equation}
\label{decdel}
\Gamma_{\Phi \rightarrow {\chi} {\bar \chi} } \sim \frac{z^{2}_{\chi}}{48 \pi} \frac{m^{3}_\Phi }{M^{2}_{P}}.
\end{equation}
(vii) Decay into singlets:  
Similar to the case of color triplets, the coupling of the modulus to the pair of singlets is given by
\begin{equation}
K=   \frac{1}{{T_B+ {\bar T}_B}}\left( {\cal N} {{\cal N}^{\dagger}} + {\bar {\cal N}} {\bar  {\cal N} }^{\dagger}\right)+  \left( \frac{z_ {\cal N} {\cal N} {\bar  {\cal N}}}{{T_B+ {\bar T}_B}} + h.c.\right),
\end{equation}
 where $z_{\cal N}$ is an undetermined constant.  After canonical normalization of the volume modulus, the interaction term is given by
\begin{equation}
{\cal L}= \frac{1}{2} \sqrt{\frac{2}{3}} \left ({\partial}^{2}  \Phi^{\dagger} \right) {\cal N} {\bar  {\cal N}}.
\end{equation}
The decay width for the modulus decay into singlets is given by
\begin{equation}
\label{decN}
\Gamma_{\Phi \rightarrow {\cal N} {\bar  {\cal N}} } \sim \frac{z^{2}_ {\cal N}}{48 \pi} \frac{m^{3}_\Phi }{M^{2}_{P}}.
\end{equation}
Since $m_{\rm soft}, m_{\rm gaugino} << m_{\Phi}$, it is clear from equation (\ref{decysc}) and (\ref{decyferm}) that  couplings of the modulus into (s)particles as well as gauginos are suppressed. The modulus can dominantly decay into any of the pair of Higgs, axions, color triplets $\chi$ and singlets ${\cal N}$ depending on the value of the coefficients of the interaction coupling of the modulus to the same. In the next subsection, we show that the coefficient of the interaction term coupling the modulus to the pair of color triplets $\chi$ and singlets ${\cal N}$, e.g. $z_{\chi}$ and $z_N$, can be constrained based on the masses of these particles.
 
For  ${\Gamma_\Phi} \sim m^{3}_{\phi}/{M^{2}_P}$, the reheating temperature after decay of the modulus is given by $ T_R=\frac{1}{{g_*}^{1/4}} \sqrt{\Gamma_{\Phi} M_P}$.  For the lightest modulus mass $m_{\phi} \sim  M_{P}/ {\cal V}^{3/2} \sim 5 \times 10^{6} \gev$ and taking ${\cal V}\sim10^7$ and ${g_*}\sim {\cal O}(100)$, we obtain $ T_R \sim  {\cal O}(1) \gev $.
\subsection{ Constraints on the modulus interaction to color triplet (singlet) superfields}
  In ${\cal N}=1$ supergravity limit of any superstring model, the effective mass term of any matter superfield obtains contribution from both superpotential as well as K\"{a}hler potential, and the values of both supersymmetric mass term and soft SUSY breaking parameter  corresponding to any matter superfield depend on the interaction of the hidden sector field  with the  matter superfields \cite{Brignole:1997dp}. Hence, one can naively expect that the coefficient of the interaction coupling of the hidden sector field to the matter superfields can be constrained depending on the mass of the matter superfields. We explicitly describe this situation in the context of sequestered large volume compactification model below.

The soft terms as well as supersymmetric mass terms for the MSSM have  already been calculated in the context of sequestered LVS model in Refs. \cite{Blumenhagen:2009gk,Aparicio:2014wxa}. Similar to the MSSM superfields, the soft SUSY mass terms of additional superfields under consideration will also depend on their interaction with moduli (hidden sector fields). Including additional color triplet superfields $({\chi}, {\bar \chi})$ and singlet superfields $({\cal N},{\bar {\cal N}})$, the matter superpotential is given by
 \begin{eqnarray}{\label{eq:sup}}
&& W_{\rm matter}=\mu(\Phi) H_{u} H_{d} +\frac{1}{6} Y_{ijk}\left(\Phi \right) C^{i}C^{j}C^{k} \nonumber\\
&& +M_{\chi} (\Phi) \chi {\bar \chi} + M_{N }\left(\Phi\right) {\cal N}  {\bar {\cal N}} + \kappa_{i} (\Phi) {\cal N}  \chi u^{c}_{i}\nonumber\\
&& + \kappa^{\prime}_{ij} (\Phi) \chi d_i d_j  + \kappa^{\prime \prime}_{ij} (\Phi) \chi \nu_i \nu_j +\cdots \; .
\end{eqnarray}
The K\"{a}hler potential corresponding to matter superfields is given by
\begin{eqnarray}
 && {\hskip -0.3in} K_{\rm matter}= {\tilde K}_{i}({\Phi, {\bar \Phi}}) \left| C^{i}\right|^2 +   {\tilde K_{\chi}} (\Phi, {\bar \Phi}) \left|{\chi} \right|^2  + {\tilde K_{\cal N}} (\Phi, {\bar \Phi})   \left|{\cal N} \right|^2  \nonumber\\
 &&  {\hskip -0.3in}  + \Bigl( Z_h(\Phi, {\bar \Phi})   H_{u} H_{d} + Z_{\chi} (\Phi, {\bar \Phi})   {\chi}{\bar \chi}  + Z_{\cal N} (\Phi, {\bar \Phi})   {\cal N}  {\bar {\cal N} }+ h.c \Bigr), \nonumber\\
 \end{eqnarray}
where  $Z_{\chi} (\Phi, {\bar \Phi}) = z_{\chi} f_{\chi}(\Phi, {\bar \Phi})$, $ Z_{\cal N}(\Phi, {\bar \Phi}) =  z_{\cal N} f_{\cal N}(\Phi, {\bar \Phi})$ and
 $C^{i}$, $i=1,2,3$ correspond to the three generations of MSSM superfields. The general expression for supersymmetric mass term associated with superfield $\chi$ is given by
 \begin{equation}
 \label{eq:effMX}
M'_{\cal  \chi}= \left( e^{\hat K/2}M_{\cal  \chi} + m_{3/2} Z_{{\cal  \chi} } - {\bar F}^{\bar I} \partial_{\bar I} Z_{{\cal  \chi} } \right) \left( {\tilde K}_{{\cal  \chi} } {\tilde K}_{{\cal {\bar \chi}} }\right)^{-1/2},
 \end{equation}
 where the index $I$ corresponds to the number of moduli. The first term in the above expression appears from the superpotential while the second and third terms appear from the K\"{a}hler potential. Similar to the higgsino mass parameter $\mu$ in the MSSM,  $M_{\chi}$ is just an input parameter which can be made to vanish under general assumptions \cite{Conlon:2006tj}.  
 Following Refs. \cite{Blumenhagen:2009gk, Aparicio:2014wxa}, mass term $M'_{\cal  \chi}$ similar to the higgsino mass parameter $\mu$ in the MSSM will be proportional to\footnote{Since second and third terms in equation (\ref{eq:effMX}) depend only on moduli, the supersymmetric mass term  remain same for all visible sector superfields except the coefficient of Giudice-Masiero term $z_i$.}
  \begin{equation}
  M'_{\chi} \sim {\cal O}(1) z_{\chi} m_{1/2} \sim  {\cal O}(1) z_{\chi}  \frac{M_P}{{\cal V}^2}.   \end{equation}
For ${\cal V} \sim{\cal O}(1) \times 10^7$, we obtain  $M'_{\chi} \sim {\cal O}(\tev)$  if one considers  $z_{\chi} = 1$.
  
 The soft SUSY breaking Lagrangian for additional superfields is given by
\begin{eqnarray}
&&  {\hskip -0.2in}  {\cal L} =  
 m^{2}_{\chi} |{\tilde \chi}|^2 + m^{2}_{\bar \chi} |{\tilde {\bar \chi}}|^2 + (B_{\chi} M_{\chi}) {\chi}{\bar \chi}+ \nonumber\\
&& {\hskip -0.2in} m^{2}_{\cal N} |{\tilde N}|^2 + m^{2}_{\bar {\cal N}} |{\tilde {\bar {\cal N}}}|^2 + (B_{\cal N} M_{\cal N}) {\cal N}{\bar {\cal N}} 
  +h.c.\; .  
  \end{eqnarray}
 The soft scalar mixing term associated with superfield $\chi$ is given by
 \begin{eqnarray}
 && {\hskip -0.3in} \left. B_{\chi} M_{ \chi} \right|_{F} =  \left({\tilde K}^{2}_{{\cal  \chi} } \right)^{-1/2} \ \left\{2 m^{2}_{3/2} Z_{{\cal  \chi} } - 2m_{3/2} {\bar F}^{\bar I} {\partial}_{\bar I} Z_{{\cal  \chi} } \right. \nonumber\\
 &&  \left. + m_{3/2} F^I \left[ \partial_I Z_{{\cal  \chi} } -   Z_{{\cal  \chi} } \partial_I {\rm ln} \left( {\tilde K}^{2}_{{\cal  \chi} } \right)\right]- \right.  
 \nonumber\\
 &&
  \left.  F^I {\bar F}^{\bar J} \left[ \partial_{I} {\partial}_{\bar J} Z_{{\cal  \chi} }- \partial_{I} Z_{{\cal  \chi} } \partial_{\bar J} {\rm ln}\left({\tilde K}^{2}_{{\cal  \chi} } \right)\right]\right\}. \nonumber\\
  \end{eqnarray}
 Following Refs. \cite{Blumenhagen:2009gk, Aparicio:2014wxa}, the  value of $ B_{\chi} M_{\cal \chi}$ (similar to the soft SUSY breaking term corresponding to the higgsino mass parameter ($B \mu$) in the MSSM) is given by
 \begin{equation}
 B_{\chi} M_{\cal \chi} \sim {\cal O}(1) z_{\chi} m^{2}_{\rm 1/2}.
 \end{equation} 
The mass matrix in the ($\chi, {\bar \chi}$) basis can be diagonalized by the transformations
\begin{eqnarray}
&& {\cal \chi} = \cos \theta \chi_{+} - \sin \theta e^{-i \phi}\chi_{-}\; , \nonumber\\
&& \bar \chi^{*} =  \sin \theta  e^{i \phi} \chi_{+} +  \cos \theta \chi_{-}\; .
\end{eqnarray}
In the diagonalized basis, the mass eigenvalues for the scalar component of ${\cal \chi}$ are given by
\begin{equation}
m^{2}_{{\cal \chi}_{\pm}}= \left|M_{\cal \chi}\right|^2 +\frac{{\tilde m}^{2}_{\cal \chi}+{\tilde m}^{2}_{\bar {\cal \chi}}}{2}+\sqrt{\frac{\left( {\tilde m}^{2}_{\cal \chi}-{\tilde m}^{2}_{\bar {\cal \chi}}\right)^2}{2} +  \left| B_{\chi} {M_{\cal \chi}} \right|^2},
\end{equation}
with
\begin{equation}
\tan{2\theta}= \frac{\left|B_{\chi} {M_{\cal \chi}}\right|}{{\tilde m}^{2}_{\cal \chi}-{\tilde m}^{2}_{\bar {\cal \chi}}}, \phi= {\rm Arg}(B_{\chi} M_{\chi}){\rm sgn} \left({{\tilde m}^{2}_{\cal \chi}-{\tilde m}^{2}_{\bar {\cal \chi}}}\right).
\end{equation}
Due to universality of the soft terms masses ${\tilde m}^{2}_{\cal \chi} = {\tilde m}^{2}_{\bar {\cal \chi}}= m^{2}_{\rm soft}$, the above equation reduces to 
\begin{equation}
m^{2}_{{\cal \chi}_{\pm}}= \left|M_{\cal \chi}\right|^2 +{\tilde m}^{2}_{\rm soft} \pm \left| B_{\cal \chi}{M_{\cal \chi}} \right|.
\end{equation}
For $z_{{\cal \chi}} = 1$, one obtains $ B_{\chi} M_{\cal \chi} =  {\tilde m}^{2}_{\rm soft}$. Hence,  the lightest mass eigenvalue of scalar component of ${\cal \chi}$ is the same as the fermion mass $M_{\cal \chi}$. Thus both scalar and fermionic components of ${\cal \chi}$ will be of ${\cal O}(\tev)$ if $z_{{\cal \chi}} =1$. 

 Similarly, we have
 \begin{eqnarray}
\label{eq:2}
&&{\hskip -0.2in} M_{\cal N} \sim {\cal O}(1) z_{\cal N}m_{\rm 1/2} \sim {\cal O}(1) z_{\cal N}\frac{M_{P}}{{\cal V}^2}.
\end{eqnarray}
For ${\cal V}= {\cal O}(1) \times 10^7$, we obtain $M_{\cal N} \sim {\cal O}(\gev)$ for  ${z}_{\cal N} = 10^{-3}$. The value of  $B_{\cal N} M_{\cal N}$ is given by
\begin{equation}
 B_{\cal N} M_{\cal N} \sim {\cal O}(1) z_{\cal N} m^{2}_{\rm 1/2}. 
 \end{equation}
 The mass eigenvalues for the scalar component of ${\cal N}$ is given by
\begin{equation}
m^{2}_{{\cal N}_{\pm}}= \left|M_{\cal N}\right|^2 +{\tilde m}^{2}_{\cal N} \pm \left| B_{\cal N} {M_{\cal N}} \right|.
\end{equation}
For ${\tilde m}^{2}_{\cal N}  \sim m^{2}_{\rm soft}$ and   $B_{\cal N} {M_{\cal N}} \sim 10^{-3}m^{2}_{\rm soft}$ , the eigenvalues of the scalar component of ${\cal N}$ will be ${\cal O}(\tev)$ even for   ${z}_{\cal N} \sim 10^{-3}$. By using equations (\ref{decdel}) and (\ref{decN}), it implies that 
\begin{equation}
\frac{\Gamma_{\Phi \rightarrow {\cal N} \bar {\cal N} }}{\Gamma_{\Phi \rightarrow {\cal \chi} {\bar {\cal \chi}} }}  \sim 10^{-6},
\end{equation}
which clearly shows that the modulus will dominantly decay into the pair of color triplet superfield $({\cal \chi},{\cal {\bar \chi}}$) as compared to the pair of singlet superfields $({\cal N},{\bar {\cal N}})$.

\section{Cogenesis Mechanism}
 In this section, we discuss the cogenesis mechanism in which both the baryon asymmetry as well as the dark matter abundance are produced via baryon number conserving decay of a singlet superfield ${\cal N}$. As mentioned earlier, we are consider the mass of the fermionic component of $({\cal N},{\bar {\cal N}})$ to be of the order of $\gev$, while the masses of   fermionic components of $({\chi}, {\bar \chi})$ to be of the order of $\tev$. The modulus decaying into any of these particles as well as the other particles will dilute the pre-existing baryon asymmetry as well as the dark matter density produced at high temperature. As a result, the baryon asymmetry and the dark matter density have to be created via some other mechanism at low temperature. It follows from the results for the decay of the modulus into different species (discussed in the previous section) that the modulus can dominantly decay into the pair of Higgs, axions and color triplets. We are interested in a scenario where the baryon asymmetry can be produced at a low temperature through the decay of ${\cal \chi }$ and in the process, one ends up with an asymmetric component of dark matter with a mass ${\cal O}(\gev)$ giving the correct relic density \footnote{Though the Higgs can further decay into the SM particles,  however it is not very relevant for explaining baryogenesis and dark matter abundance in our scenario. Similar conclusions follow for closed (open) string axions.}. 
 \subsection{Baryon asymmetry and  asymmetric dark matter}
In this subsection, we describe a mechanism of baryogenesis  where the fundamental interactions conserve baryon number, and the baryogenesis happens by generating certain amount of asymmetry in both the visible and the dark sector. We show that the decay products of the lightest eigenstate of ${\cal \chi}_{-}$( ${\cal {\bar \chi}}_{-}$) can simultaneously explain the observed baryon asymmetry and give rise to an asymmetric dark matter component, if ${\cal N}$ is light (mass ${\cal O}(\gev)$). The ratio of dark matter abundance to baryon abundance is given by
\begin{equation}
\frac{\Omega_{\rm ADM}}{\Omega_{\rm B}}=\frac{{\cal Y}_{\rm ADM}}{{\cal Y}_{\rm B}}\frac{m_{\rm ADM}}{m_B}.
\end{equation}
The cosmic coincidence ${\Omega_{\rm ADM}}/{\Omega_{\rm B}} \sim 5$ is satisfied if ${\cal Y}_{\rm ADM} \sim {{\cal Y}_{\rm B}} \sim 10^{-10}$  
for a dark matter mass $M_{\cal N}$ around 5 $\gev$. In the previous section, we have seen that modulus decays dominantly into pair of ${\cal \chi}$ and ${\cal{\bar \chi}}$ superfields. A cartoon showing the decay of the modulus into color triplet superfields which further decay into a singlet and a quark is given in Figure 1.
\begin{figure}[h]
	\includegraphics[width=1.5in, height=1.3 in]{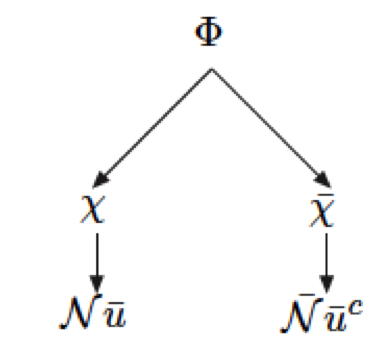}
	\caption{A cartoon showing decay of the modulus into color triplet and singlet superfields.}
	\label{fig:wn3} 
\end{figure}
Using the interaction term given in equation \ref{eq:sup}, it follows that lightest mass eigenstate of ${\cal \chi}$(${\cal {\bar \chi}}$) decays into quarks and fermionic component of the singlet superfield ${\cal N}$(${\cal {\bar N}}$). The ${\chi}$ and  ${\bar \chi}$ have baryon number assignments  $B = +2/3$ and $B = -2/3$ respectively, while ${\cal N}$ and  ${\bar {\cal N}}$  have  baryon numbers $B = +1$ and $B = -1$ respectively. Therefore, the interaction of ${\cal \chi}$ with  ${\cal N}$ and ${\bar u}$ conserves baryon number. The scalar and fermionic components of ${\chi}$ and ${\bar \chi}$ have $R$-parity assignments +1 and -1 respectively, while the scalar and fermionic components of ${\cal N}$ and ${\bar {\cal N}}$ have R-parity assignments -1 and +1 respectively. Now $R$-parity conservation implies that the fermionic component of ${\cal N}$(${\cal {\bar N}}$) will further not decay into the SM particles. Consequently, the fermionic component of ${\cal N}$ superfield produced during decay of  $\tilde{{\cal N}_{-}}$ can be considered as a stable asymmetric dark matter particle. Thus, it follows that the decay of a scalar component of ${\cal N}$ generates an equal and opposite amount of baryon asymmetry in the visible sector and the dark sector i.e.
 \begin{equation}
 \label{eq:Btot}
 {\cal Y}_{B}+{\cal Y}_{DM}=0 \Rightarrow {\cal Y}_{DM} = -{\cal Y}_{B}.
\end{equation}
 \begin{figure}[h]
	\includegraphics[width=3.5in]{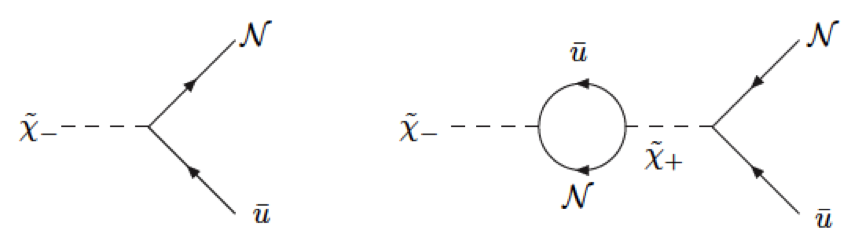}
	\caption{Feynman diagrams contributing to baryon asymmetry and dark matter asymmetry via baryon number conserved interaction vertex.}
	\label{fig:wn3} 
\end{figure}

Let us first calculate the baryon asymmetry generated in the visible sector. We begin by considering a single generation of singlet superfield ${\cal N}$. The CP violation arises due to the soft SUSY breaking terms. The degeneracy between the two states belonging to the supermultiplet of a same generation can be removed by including the SUSY breaking effects and the CP violation occurs due to the interference between the two states of a single generation \cite{Grossman:2003jv,D'Ambrosio:2003wy}, as compared to conventional baryogenesis mechanism where at least two generations are required for CP violation. The SUSY breaking Lagrangian involving ${\cal \chi}$ and ${\bar \chi}$ superfields is given by
 \begin{eqnarray}
 &&  {\cal L}_{\rm soft}= m^{2}_{ij} {{\tilde {\chi}}_i}^{\dagger} {\tilde \chi}_j+m^{2}_{ij}{\tilde {\bar {\chi}}^{\dagger}_i} {\tilde {\bar \chi}}_j+  B_{\chi ij} M_{\chi ij}  {\tilde \chi_i}  {\tilde {\bar {\chi_j}}} \nonumber\\
 && + A_{ijk} \kappa_{ijk} {\tilde {\cal \chi}_i}  {\tilde {\cal N}_j} {\tilde u}^{*}_k+h.c.\; ,
 \end{eqnarray} 
 where indices $i,j,k$ correspond to the different generations of the  particles. The evolution of system  governing ${\cal \chi}-{\cal \chi}^{\dagger}$ mixing is given by \cite{Grossman:2003jv,D'Ambrosio:2003wy}
\begin{equation}
\langle {\tilde \chi}|{\cal H}|{\tilde \chi}^{\dagger}\rangle = M_{\chi(12)} - \frac{i}{2} \Gamma_{\chi(12)}.
\end{equation}
This induces a mass difference $\Delta M_{\chi} $ and a decay width difference $\Delta\Gamma $ between two states given by
\begin{equation}
 |{\tilde \chi}_{\pm}\rangle = p |{\tilde \chi}\rangle \pm  q |{\tilde \chi}^{\dagger}\rangle, \; \; R=|q/p|.
\end{equation}
The tree level and the one-loop diagrams  responsible for generating CP violation are shown in Figure 2. The interaction Lagrangian in the mass basis $({\tilde \chi}_{+}, {\tilde \chi}_{-})$ is given by,
\begin{eqnarray}
&& -{\cal L}_{\rm int}= \frac{\kappa}{\sqrt{2}}{\tilde \chi}_{+} {\cal N} {\bar u} + \frac{\kappa}{\sqrt{2}}{\tilde \chi}_{-} {\cal N}  {\bar u} +h.c. \; .
\end{eqnarray}
Now, if we start with equal densities of ${\tilde \chi}$ and ${\tilde \chi}^{\dagger}$ at $t=0$, then after some time $t$ the states evolve into 
\begin{eqnarray}
&& {\tilde \chi}(t)= f_{+}(t) {\tilde \chi}(0) + \frac{q}{p} f_{-}(t) {\tilde \chi}^{\dagger}(0), \nonumber\\
&&  {\tilde \chi}^{\dagger}(t)= \frac{q}{p} f_{-}(t) {\tilde \chi}(0) +  f_{+}(t) {\tilde \chi}^{\dagger}(0),
\end{eqnarray}
where 
\begin{eqnarray}
 && f_{+}(t) = e^{-iM_{\chi} t} e^{-\Gamma_{\chi} t/2} {\rm cos}({\Delta M}_{\chi} t/2), \nonumber\\
 &&  f_{-}(t) = i e^{-iM_{\chi} t} e^{-\Gamma_{\chi} t/2} {\rm sin}({\Delta M}_{\chi}  t/2).
\end{eqnarray}
Now the excess of ${\cal N}, {\bar u}$ over ${\cal N}^{c}, {\bar u}^c$ is given by
\begin{widetext}
\begin{equation}
\label{asymmetry}
\epsilon_{\cal \chi}= \frac{\int^{\infty}_{0} dt\left[ \Gamma\left({\tilde \chi(t)} \rightarrow {\cal N} {\bar u} \right)+\Gamma({\tilde \chi(t)}^{\dagger} \rightarrow {\cal N} {\bar u}) - \Gamma\left({\tilde \chi(t)} \rightarrow{\bar  {\cal N}} {\bar u}^c\right) - \Gamma({\tilde \chi(t)}^{\dagger} \rightarrow{\bar  {\cal N}} {\bar u}^{c})\right]} {\int^{\infty}_{0} dt\left[ \Gamma({\tilde \chi(t)} \rightarrow {\cal N} {\bar u})+\Gamma({\tilde \chi(t)}^{\dagger} \rightarrow {\cal N} {\bar u}) + \Gamma({\tilde \chi(t)} \rightarrow{\bar  {\cal N}} {\bar u}^c) + \Gamma({\tilde \chi(t)}^{\dagger} \rightarrow {\bar  {\cal N}} {\bar u}^c)\right]}.
\end{equation}
\end{widetext}
 The decay width for the decay modes of ${\tilde \chi(t)}$ and ${\tilde \chi(t)}^{\dagger}$ is given by\footnote{Since we are working with a single generation, the direct CP violation \cite{Nir:1992uv} can be neglected. Therefore, the amplitude of the decay process (${\tilde \chi(t)} \rightarrow {\cal N} u^c$) and its CP conjugate state are the same.}
 \begin{eqnarray}
 && {\hskip -0.2in} \Gamma({\tilde \chi(t)} \rightarrow {\cal N} {\bar u}) = \Gamma({\tilde \chi(t)}^{\dagger} \rightarrow{\bar  {\cal N} } \bar {u^{c}_i})= X_1 |M_{u {\cal N} }|^2 f_{+}(t),  \nonumber\\
 &&  {\hskip -0.2in} \Gamma({\tilde \chi(t)}^{\dagger} \rightarrow {\cal N}   {\bar u}) = X_1 |M_{u {\cal N} }|^2 R^{-2} f_{-}(t),  \nonumber\\
 &&  {\hskip -0.2in} \Gamma({\tilde \chi(t)} \rightarrow{\bar  {\cal N} } {\bar u}^c)=X_1 |M_{u {\cal N} }|^2 R^{+2} f_{-}(t),
  \end{eqnarray}
where $X_1$ is the normalization factor and $M_{u  {\cal N} }$ is the amplitude of the decay mode considered. Incorporating these expressions into equation (\ref{asymmetry}), the asymmetry parameter is given by  \cite{Grossman:2003jv,D'Ambrosio:2003wy}
\begin{equation}
\epsilon_{\cal \chi}= \frac{1}{2}\left( \left|\frac{q}{p}\right|^2 -  \left|\frac{p}{q}\right|^2\right) \frac{\int^{\infty}_{0} dt \left|f_{-}\right|^2}  {\int^{\infty}_{0} dt \left(\left|f_{+}\right|^2 + \left|f_{-}\right|^2\right)},
\end{equation}
with 
 \begin{equation}
 \left|\frac{q}{p}\right|^2 -  \left|\frac{q}{p}\right|^2  \sim  {\rm Im} \frac{\Gamma_{\chi(12)}}{M_{\chi(12)}} =  \frac{\Gamma_\chi \  {\rm Im}{A}}{M_\chi . \ B_\chi M_\chi},
 \end{equation}
 and 
 \begin{equation}
\frac{\int^{\infty}_{0} dt \left|f_{-}\right|^2}  {\int^{\infty}_{0} dt \left(\left|f_{+}\right|^2 + \left|f_{-}\right|^2\right)}= \frac{ (\Delta M_\chi)^2}{2(\Gamma^2 +({\Delta M_\chi})^2)}.
\end{equation}
The decay width for ${\cal \chi}_{-}$ decaying into ${\cal N}{\bar u}$ is given by $\Gamma_{\cal \chi}= \frac{\kappa \kappa^{\dagger}}{4 \pi} M_{\chi}$. The asymmetry is given by
\begin{equation}
{\epsilon}_{B}= \frac{\Gamma_\chi}{\Gamma^{2}_{\chi} + B^{2}_{\chi} M^{2}_{\chi}}\frac{  B_{\chi} M_{\chi} \ {\rm Im} A}{2 M_\chi}. 
\end{equation}
Using equation (\ref{eq:scales}) we have $B_\chi M_\chi = M_\chi \sim {\cal O}(\tev)$,  and $\kappa \lesssim 1$, implying $\Gamma_\chi < B_\chi M_\chi$. In the small $\kappa$ limit, the above equation reduces to
\begin{equation}
{\epsilon}_{B}= \frac{\kappa   \ {\rm Im} A}{ 8 \pi}  \frac{M_\chi}{B_\chi M_\chi}.
\end{equation}
The baryon asymmetry generated from the modulus decay is given by
\begin{equation}
\label{etaB}
{\cal Y}_{B} = \frac{n_B - n_{\bar B}}{n_{\gamma}} = {\cal Y}_{\cal \chi} \epsilon_{B}.
\end{equation}
 Since the Yukawa couplings do not depend on the compactification volume in this model, we choose $\kappa \sim 10^{-2}$. For ${\cal A}_{ijk} \equiv {\kappa}_{ijk} A_{ijk}\sim {\cal O}(\tev)$, we get $\epsilon_{B} =10^{-3}$.  
 
 Now using equation (\ref{eq:YX}), the value of ${\cal Y}_{\cal \chi}$ is given by
\begin{equation}
Y_{\chi}= Y_{\Phi} {\rm Br}_{\chi}= \frac{ 3 T_R}{4 M_\Phi} {\rm Br}_{\chi}
\end{equation}
For ${\rm Br}_{\chi}=1, T_R \sim {\cal O}(\gev)$ and $m_{\Phi} \sim 5 \times 10^6 \gev$, we get $Y_{\chi} \sim 10^{-7}$. Finally, from equation (\ref{etaB}) the baryon asymmetry is given by ${\cal Y}_{B} \sim 10^{-3-7} = 10^{-10}$, which is the observed baryon asymmetry of the universe. From equation (\ref{eq:Btot}), the asymmetry in the dark matter sector is also given by ${\cal Y}_{\rm DM} \equiv {\cal Y}_{\cal N} \sim 10^{-10}$.
 \vskip 0.1in
\subsection{Annihilation of the symmetric dark matter component} The requirement that an overall dark matter abundance is asymmetric follows from the condition that the symmetric component of dark matter should get depleted. For this to happen, the annihilation cross section of the symmetric component should be higher than the value of the cross section at freeze out temperature \cite{Baer:2014eja}. The freeze out temperature can be estimated by the rule of thumb  given by
\begin{equation}
\label{eq:sigmav}
\Gamma =n \langle \sigma \left|v\right| \rangle = H.
\end{equation} 
similar to the case of WIMP dark matter. However, in this case the reference temperature is the reheating temperature after the decay of the modulus. The Hubble expansion rate is given by $H= \frac{T^{2}_R}{M_P}$. The decay of the modulus produces a baryon number symmetric component of dark matter with a density given by $n= {\rm Br_{\chi}}\frac{3 T_R}{4 m_{\Phi}} \times s $, where s is the entropy density. Using equation (\ref{eq:sigmav}) for $T_R= {\cal O}(\gev)$ and $s= \frac{2 \pi^2}{45} g_{*} T^{3}_R$ , the thermally averaged cross section is given by
\begin{equation}
 \langle \sigma \left|v\right| \rangle =  \frac{30}{ \pi^2 g_{*}} \frac{  m_{\Phi}}{T^{2}_R M_P}.
 \end{equation}
  For the modulus mass $m_{\Phi}= 5 \times 10^6 \gev$ and $g_{*}= {\cal O}(100)$, we get $\langle \sigma \left|v\right| \rangle_{\rm freeze~out} \sim {\cal O}(10^{-14}$) $\gev^{-2}$.

The annihilation of ${\cal N}$ and ${\cal N}^c$ can be mediated through electroweak neutral $Z$ boson. The cross section for ${\cal N} {\cal N}^c \rightarrow Z \rightarrow f {\bar f}$ is given by
\begin{equation}
\langle \sigma \left|v\right| \rangle \sim \frac{1}{4 \pi} \frac{ g^{\prime 2} g^2 M^{2}_{\cal N}}{M^{4}_{Z}},
\end{equation}
where $g^{\prime}$ corresponds to the gauge coupling of neutral $Z$ boson to the pair of singlet superfields $({\cal N}, {\bar{\cal N}})$, $g$ corresponds to the gauge coupling of $Z$ boson to a pair of fermions and $M_Z$ is the mass of $Z$ boson. For $g^{\prime} \sim 1$, $M_{\cal N}=5 \gev$, we get $\langle \sigma \left|v\right| \rangle \sim 10^{-8} \gev^{-2}$, which is higher than the freeze out cross section calculated above. Thus, the symmetric component of the abundance of singlet fermion ${\cal N}$ gets annihilated and the left-over asymmetric relic abundance gives the correct dark matter relic abundance.
\section{Concluding remarks}
We have discussed a cogenesis mechanism unifying the generation of both baryon asymmetry of the universe and dark matter abundance in a model which can be obtained from a low energy limit of type IIB LVS string model. The decay of the lightest modulus (generically present in string models) dilutes the pre-existing baryon asymmetry and dark matter abundance of the universe.  To avoid the cosmological moduli problem (CMP) one requires a very heavy modulus, which decays post inflation to give a low reheating temperature. Therefore one needs to consider a post reheating mechanism for generating the baryon asymmetry as well as the dark matter abundance. In this work, we show that  both the baryon asymmetry and the non-thermal dark matter abundance can be generated simultaneously from the decay of a pair of color triplets produced after reheating. We consider the mass of the fermionic component of the pair of color triplets and singlets to be $1 \tev$ and $5 \gev$ respectively. We demonstrate that in the context of ${\cal N}=1$ supergravity, the interaction coupling of the modulus to the pair of singlet superfields as well as the colored superfields can be constrained depending upon the masses of the same. We find that the branching ratio of the modulus decaying into the pair of additional singlets is suppressed by a factor of $10^{-6}$ as compared to the decay of the modulus into the pair of color triplets. Therefore, we conclude that the modulus will dominantly decay into pair of color triplets. The lightest eigenstate of scalar component of the color triplet further decays into singlet fermion and up type quark. Imposition of $R$-parity conservation ensures that the singlet fermion does not further decay into the SM particles and therefore it can be considered as a stable dark matter candidate. The CP asymmetry is generated via the interference of tree level and one loop diagrams for the decay of color triplets in the presence of soft SUSY breaking terms. We find that it is possible to obtain the observed baryon asymmetry of the universe and the asymmetric dark matter abundance by considering dark matter mass around 5 $\gev$, and the cosmic coincidence is natural in this scenario. Thus, if $\tev$ scale colored fields are found at the LHC, it will have a very profound consequences for explaining the observed baryon asymmetry of the universe, the dark matter abundance and the cosmic coincidence.   
\section*{Acknowledgments}
 MD and CH are thankful to ICTP, Trieste for financial support and hospitality where part of this work was done.  MD and CH would like to thank Wilfried Buchm\"{u}ller, Markus Luty, Subir Sarkar, Namit Mahajan and Raghavan Rangarajan for many helpful discussions. MD would also like to acknowledge Aalok Misra and Gaurav Goswami for useful discussions.

\end{document}